\documentclass[12pt,ams,a4paper,nofootinbib]{revtex4}
\usepackage{amsmath}
\usepackage{amssymb}
\usepackage{graphicx}
\usepackage[usenames,dvipsnames]{color}

\newcommand{\re}{\ref}
\newcommand{\be}{\begin{equation}}
\newcommand{\ee}{\end{equation}}

\newcommand{\la}{\label}
\newcommand{\ber}{\begin{eqnarray}}
\newcommand{\eer}{\end{eqnarray}}

\newcommand{\lcdot}{\boldsymbol{\cdot}}
\begin{document}

%\centerline{Indian J Phys {\bf 100}, 1561-1567 (2026)}

\title{On the representation-free formalism in quantum theory}

\author{
%\bf \copyright  
V.D. Efros\footnote{v.efros@mererand.com}%; ORCID: 0000-0001-9395-6308.}
}

\affiliation{
National Research Centre "Kurchatov Institute"%,
%, 123182 
%Moscow, Russia
}

\begin{abstract}

The widespread bra-ket formalism offers valuable tools for conducting representation-free
 considerations in quantum theory. However, it is not without its drawbacks. In this work, we
  discuss these drawbacks in detail and subsequently construct a new representation-free
   scheme. Similar to the bra-ket formalism, the present scheme provides essential means to
    facilitate representation-free considerations. At the same time, it is entirely free from the
     drawbacks of
the bra-ket formalism. Unlike the dual-space bra-ket formalism, the present scheme allows
 both one-space and dual-space interpretations, which is a beneficial feature.
The constructed universal scheme is very well-suited for performing representation-free practical
 calculations.

\end{abstract}

%\keywords{Quantum theory, representation-free approach, bra-ket formalism, 
%one-space and dual-space interpretations}

\pacs{03.65.-w, 03.65.Ca, 03.65.Ta}

\maketitle

%\pagebreak

%\centerline{\bf 1 INTRODUCTION}
%\hspace{1.5cm}
\section{Introduction}
In quantum theory, the representation-free approach
 is indispensable when working with general relations. It seems appropriate to 
 compare the strengths of this approach with
 those of, for example, the vector
 form of the Maxwell equations. Typically, physicists adopt the bra-ket formalism as
the representation-free approach. This formalism provides helpful tools~\cite{dir1,dir2}
for conducting practical calculations.  However, it is not satisfactory to a substantial
degree, and various authors drew attention to its drawbacks~\cite{gie,ja,gr,wei,si}. 

Besides those drawbacks, 
the bra-ket formalism is a dual-space scheme, which presents a substantial limitation. 
A more straightforward and natural one-space formulation would be preferable in many cases.
Mathematicians typically employ
the formulation, listed below, of such a type in studies both on abstract Hilbert space and
 quantum theory. Weinberg~\cite{wei} also 
utilized  this formulation.
This formulation does not have drawbacks of the bra-ket formalism but
it lacks
helpful tools that the bra-ket formalism provides. Furthermore, 
while this one-space formulation does not include bra vectors,  they
are still required for various applications, and   
the dual-space approach incorporating bra vectors 
is needed 
for noteworthy purposes, see, e.g., Refs.~\mbox{\cite{gie,gel,bo,oh,oh1,gan,ott}}. 
Until now, the bra-ket  formulation and this one-space 
 formulation have been the only ones used in 
the literature.

In Sec.~3, we construct a convenient and universal representation-free scheme that, 
in contrast to the aforementioned formulations, allows both
the one-space and dual-space interpretations. Furthermore, unlike the  aforementioned
 mathematically oriented formulation, the present scheme provides tools 
that facilitate practical calculations, of the type of tools inherent in the bra-ket formalism.
With all that, the present formulation is entirely free from 
the drawbacks of the bra-ket formalism.

Before this,  we address the strengths and drawbacks of the bra-ket 
formalism in Sec.~2. In particular, the original version~\cite{dir1} of this formalism
is critically analyzed in detail,
which, to our knowledge, has not been done previously. 

One may note that any representation-free formulation of quantum theory should be
 equivalent to others, provided that they all are correct and complete. However, this by no  
means implies that they would be equally suitable for practical calculations.

%\centerline{\bf 2 BRA-KET FORMALISM}
\section{Bra-ket formalism}

%\subsection{Drawbacks}

%One of them is that existing matrix elements  can be represented by 
%the corresponding bra-ket expressions like $\langle u|O|v\rangle$
%not in all the cases. Before discussing this we shall examine the bra-ket
%formalism. The considerations of this section will also be of use in Sec.4.
%The bra-ket scheme as discussed in this section will be compared with the scheme
%of the present work in Secs. 3 and 4. 

In the first part of this section,
 we will analyze the original version of the
bra-ket formalism given in the classical book~\cite{dir1}. There exists a later version of this
 formalism discussed in
Ref.~\cite{gie}, which we will describe next.
Following that, we will outline the appealing properties of the bra-ket formalism, 
and discuss the drawbacks of its later version.

The present analysis of the original version of the bra-ket formalism is needful, considering that
this version
 is still followed either in full, as in Ref.~\cite{mes}, or "in part," as in Ref.~\cite{co}. A clear
 distinction between the two versions has not been made in the literature.\footnote{In
  Ref~\cite{co}, the reasoning from Ref.~\cite{dir1} is reproduced regarding the action of
   operators on bra vectors and the definition of adjoint operators. However, the definition of bra
    vectors themselves  adopted there, which is the same as in Subsection B, differs from that in 
Ref.~\cite{dir1} and is incompatible with the aforementioned reasoning.} In the considerations
 of this section, state vectors are assumed to have finite norms.

\subsection{Original version}
%\centerline{2.1\quad  Original version}

According to  Secs.~6-8 of Ref.~\cite{dir1}, the bra-ket scheme
is constructed through the following main steps.

1. One considers linear functionals~$F$ that depend on state 
 vectors~$|v\rangle$.  
 One assumes that there exists an anti-linear one-to-one 
 correspondence between the set of these functionals 
 and the set of state vectors themselves.
 Denoting state vectors   as~$|u\rangle$ in this context, 
 this correspondence can be expressed as 
 $F\Leftrightarrow|u\rangle$.
 To designate the  functionals $F$ based on this 
 correspondence, 
 the symbol~$\langle u|$ is introduced: $F\Rightarrow\langle u|$.
 Correspondingly, the   values of the functionals $F(|v\rangle)$
 are denoted as $\langle u|v\rangle\equiv\langle u|(|v\rangle)$. 
  The quantities~$\langle u|v\rangle$, referred to as scalar products,
 are thus
 linear in~$|v\rangle$ and anti-linear 
 in~$|u\rangle$.  
 It is assumed that
   $\langle u|v\rangle^*=\langle v|u\rangle$ which is consistent with the above. 
   It is also assumed that 
   $\langle v|v\rangle\ge0$ for any~$|v\rangle$. 
 
The functionals~$\langle u|$ are referred to as bra vectors. State
 vectors are called ket vectors. The ket and bra vectors that form a 
 $\langle u|\Leftrightarrow|u\rangle$ pair 
 are termed conjugates of each other.
  
 2. The action of linear operators on bra vectors is defined. 
Let $O$ be a linear operator that
acts on ket vectors. To give meaning to its action on bra vectors, the  scalar
products of the form 
$\langle u|\bigl( O|v\rangle\bigr)$ are considered.  These scalar products
are viewed,
for a given  bra vector~$\langle u|$
and  a given operator $O$, as
 the values of a linear functional that depends on 
 states $|v\rangle$.
 According to the definition above,  this
 functional is some bra vector. 
 Denoting this bra vector as $\langle w|$, 
one then may express the values of this functional as $\langle w|v\rangle$, as above. That is
 $\langle u|\bigl( O|v\rangle\bigr)=\langle w|v\rangle$.

Further, for
 a given  operator $O$, the family of the $\langle u|\bigl( O|v\rangle\bigr)$ functionals 
specified by the bra vectors~$\langle u|$ 
   is considered. 
 The  operator $O$, acting in the space of ket vectors,  uniquely defines
 an operator in the space of bra vectors that
 realizes 
 the mapping $\langle u|\Rightarrow\langle w|$ with the above definition of $\langle w|$. 
 Then, one designates the latter operator with the same 
 symbol~$O$: $\langle w|=\langle u|O$. 
 With this definition, one has
 \be\bigl(\langle u|O\bigr)|v\rangle=
  \langle u|\bigl( O|v\rangle\bigr)\stackrel{def}=\langle u|O|v\rangle.
 \la{1}\ee
 The operator in the space of
  bra vectors, thus introduced,  is linear. 

3.
 The  adjoint operator $O^\dag$  of a linear operator $O$ is defined. 
 The definition consists of the statement that
 the ket vector $O^\dag |u\rangle$ is conjugate of the bra vector
 $\langle u|O$ defined above. That is,
 \be\langle v| \bigl(O^\dag|u\rangle\bigr)\stackrel{def}=
 \left[\bigl(\langle u| O\bigr)|v\rangle\right]^*.\la{ad0}\ee 
 Using this definition along with Eq. (\re{1}) applied to both  $O^\dag$ and $O$ operators,
 one then gets the relation
 \be \langle v| \bigl(O^\dag|u\rangle\bigr)=\bigl(\langle v| O^\dag\bigr)|u\rangle
 \stackrel{def}=
 \langle v| O^\dag|u\rangle=\langle u| O|v\rangle^*.\la{2}\ee

Let us discuss the listed reasoning. First, consider step 1 above. 
We shall need the following definition. A linear functional $F$ is said to 
be unbounded if 
states  $|v\rangle$
can be found  such that 
the quantity
$\langle v|v\rangle^{-1/2}|F(|v\rangle)|$ exceeds any given constant. Otherwise, $F$
is said to be
bounded. Jauch~\cite{ja} has noted that it is incorrect to define bra vectors
in terms of unbounded linear functionals. Here, we shall give a simple explanation of this. 
%which is very short. 
In our case, $F=\langle u|$, and $F(|v\rangle)=\langle u|v\rangle$. 
According to the Schwarz inequality (that follows merely from the definition of the 
scalar product, see, e.g., \cite{reed}),
%\footnote{The Schwarz inequality follows merely
%from the fact that norms of vectors on $\cal{H}$ are non-negative. For obtaining 
%this inequality, it is sufficient
%to calculate the norm of the vector $|u\rangle-c|v\rangle$ with 
%$$c=\langle v|u\rangle/\langle v|v\rangle$.}  
one has
\[
\langle v|v\rangle^{-1/2}|\langle u|v\rangle|\le\langle u|u\rangle^{1/2}.\]
The right-hand side of this inequality is a constant independent of $|v\rangle$.
Thus, the linear functional $\langle u|$ cannot be unbounded. That is, not any linear
functional $F=\langle u|$ 
can be put into correspondence with a state vector $|u\rangle$ in a way that 
$F(|v\rangle)=\langle u|v\rangle$ possesses the prescribed properties of the scalar product.
In other words, the 
definition of bra vectors in the  reasoning listed above
is incompatible with the natural properties
of the scalar product that are adopted within that reasoning.
Thus, in contradiction to that reasoning,
not every linear functionals $F(|v\rangle)$ possesses the prescribed properties
of bra vectors.  Unbounded $F(|v\rangle)$ do not exhibit these properties. 
(Certain bounded~$F(|v\rangle)$ 
are also not bra vectors. These include some of $F(|v\rangle)$ involved in the step 2
of the aforementioned reasoning, as can be concluded from the considerations that follow.)

Jauch~\cite{ja} has suggested associating bra vectors  exclusively with
bounded linear functionals (see also below).
We contend, however, that this is insufficient to cure the formalism. 
The point is as follows.  
The reasoning in step~2 above, which addresses
the action of operators on bra vectors,
 was based on the assertion that the linear functionals~$F$ of the form
 \mbox{$F(|v\rangle)=\langle u|\bigl(O|v\rangle\bigr)$} can be considered bra
 vectors. If one denotes these bra vectors as $\langle w|$, the 
 relation expressing this assertion can be written as
 \mbox{$\langle u|\bigl(O|v\rangle\bigr)=\langle w|v\rangle$}. 
 (\mbox{$\langle w|\Rightarrow\langle u|O$}.) 
 Let us show that this relation is, in general,  invalid since the state $|w\rangle$ may not exist. 
  
 For example, let us consider a self-adjoint operator $O$ with a discrete 
 unbounded spectrum, specifically focusing on the single-particle problem 
and the operator ${\hat l}^2$
of the square of orbital momentum. 
Let $|nlm\rangle$ be eigenstates of the  ${\hat l}^2$ operator, 
\mbox{${\hat l}^2|nlm\rangle=l(l+1)|nlm\rangle$}. 
Normalize them to unity.
In the relation  we are discussing, with $O={\hat l}^2$,
 let us 
 take $|nlm\rangle$ 
 as~$|v\rangle$. 
 Then this relation transforms to \mbox{$l(l+1) u^*(nlm)=w^*(nlm)$} 
 where \mbox{$u(nlm)=\langle nlm|u\rangle$} and  
 \mbox{$w(nlm)=\langle nlm|w\rangle$} are the 
 wave functions of the  states $|u\rangle$ and  $|w\rangle$ 
 in the~$nlm$ representation.

Suppose, for example, that $|u\rangle$ is a state with fixed
 $n$ and $m$ values. Assuming that the  state~$|w\rangle$
entering the relation under discussion exists, we can  calculate the
norm of the state $|w\rangle-\sum_{l\ge|m|}^{l_{max}}w(nlm)|nlm\rangle$. This leads to
the usual-type 
inequality: \mbox{$\sum_{l\ge|m|}^{l_{max}}|w(nlm)|^2\le \langle w|w \rangle$.}
  From this inequality, it follows that the series 
 \mbox{$\sum_{l\ge|m|}^\infty [l(l+1)]^2|u(nlm)|^2$}
  must be convergent
 for the state $|w\rangle$ to exist. However, it is clear that 
 this condition is not universally satisfied, as the only requirement 
 regarding the state $|u\rangle$
  is that the series
 $\sum_{l\ge|m|}^\infty |u(nlm)|^2$ is convergent.  Thus, for certain  states $|u\rangle$
 and  operators $O$ of interest, the equality of the form
 \mbox{$\langle u|\bigl(O|v\rangle\bigr)=\langle w|v\rangle$} cannot hold.
   Consequently, the reasoning 
 of step~2  we are discussing is not valid. This implies that, in addition to bra vectors
  themselves,
  the action of operators on bra vectors also cannot be defined according to this reasoning.

Let us also note that, even with the sum
\mbox{$\sum_{l\ge|m|}^\infty|u(nlm)|^2$} being convergent, the behavior
\mbox{$l(l+1)|u(nlm)|\rightarrow\infty$} at $l\rightarrow\infty$ 
is still possible. At least for states~$|u\rangle$ exhibiting this behavior,
the functionals
\mbox{$F(|v\rangle)=\langle u|\bigl({\hat l}^2|v\rangle\bigr)$} are unbounded
and cannot be presented in the  form $\langle w|v\rangle$ for this reason alone.
 
And, as for the definition of the adjoint operator, which corresponds to step~3 of the 
 reasoning under discussion, this definition
 involves objects like $\langle u|O$. However, as argued
 above, such objects are, in fact, not properly defined in the previous step of 
 the reasoning. Therefore, 
 the definition of the adjoint operator being discussed also cannot be adopted.
 A different definition is listed below.   

\subsection{Revised version}
 
% Passing to the corrected version of the bra-ket scheme we shall 
 %employ the 
 %conventional mathematically-oriented notation like $(u,v)$ 
 % for the quantum mechanical
 %scalar product  where both  $u$ and $v$ are vectors in $\cal H$. 
%The above discussed point that bra vectors
% cannot be identified with unbounded linear functionals
% was mentioned in Ref.~\cite{ja}. 

The corrected definition of bra vectors proceeds from the usual mathematically oriented 
expression $(u,v)$
for the scalar product where both $u$ and $v$ 
are state vectors belonging to
 the Hilbert space~$\cal H$.  ($(v,u)=(u,v)^*$, and $(v,v)\ge0$ for any~$v$.)
 Let us use the symbol $G_u$ to denote special linear functionals, termed covectors,
 such that 
 \mbox{$G_u(v)=(u,v)$}. 
Jauch~\cite{ja} has suggested identifying bra vectors 
 with covectors (or bounded linear functionals\footnote{According to the
  Riesz theorem, see, e.g., \cite{reed},
 any 
 bounded linear functional defined on 
  the entire space $\cal H$ is just of the form
  $(u,v)$, where $u$ is a  given {\it unique} vector. Note that the linear functionals of
  the form $\langle u|\bigl(O|v\rangle\bigr)$ that we considered above were defined on
   subspaces of vectors $|v\rangle$ which
  are smaller than $\cal H$.}), which is adopted below.
  Thus, from now on, $\langle u|$ will denote the covector $G_u$. 
 The equality \mbox{$\langle u|v\rangle\equiv\langle u|(|v\rangle)=(u,v)$}
 is thus adopted.

Let us now
 list  the revised definitions of the action of an operator on a bra vector and of
 the
 adjoint operator, see, e.g., Ref.~\cite{gie}.
   In contrast to the previously listed version of the
 bra-ket formalism, now the 
 adjoint operator is introduced first.
 It is defined in accordance with the relation
   \be \langle v|\bigl(O^\dag|u\rangle\bigr)
=\left[\langle u|\bigl(O|v\rangle\bigr)\right]^*\la{ad}\ee 
  Eq. (\re{ad}) is different from Eq.~(\re{ad0}).  

The revised definition for the action of an $O$ operator on a covector 
is as follows: the covector  $\langle u|O$ is defined as the conjugate
 of the vector $O^\dag|u\rangle$
where the  operator~$O^\dag$  is given by Eq.~(\re{ad}). 
With this definition, the equality~(\re{1}) is valid due to 
the relation~(\re{ad}).
The definition 
for the action of an $O^\dag$ operator on a covector  is that 
the  covector  $\langle v|O^\dag$ is the conjugate
 of the vector $O|v\rangle$. With this definition, the equalities~(\re{2})
 are valid due to 
the relation~(\re{ad}).

%Regarding these definitions, we want to point out the following. 
%Both the original and the revised 
% versions of the bra-ket formalism include the statement that the vector
% $O^\dag|u\rangle$  and the covector $\langle u|O$  are conjugate of each other. 
% However, the meaning of this statement differs between the two versions. 
% In the revised version, this statement is the definition of 
% the $\langle u|O$ object in terms of the $O^\dag|u\rangle$ object.
 %since this version defines the
 %latter object in advance.
 %The corrected version of the scheme defines the
 %$O^\dag|u\rangle$ object in advance, and this statement is the definition of 
 %the $\langle u|O$ object. 
% Consequently, this statement
%  is assumed to be valid only for those state vectors~$|u\rangle$ that belong
%  to the domain of definition of the  operator $O^\dag$, cf. below. 
%  In cases of interest, this domain
%  is often less than
%  the entire space~$\cal H$. In contrast,
% in the original version of the formalism, this statement is considered to be
% the definition of
% the $O^\dag|u\rangle$ object in terms of the $\langle u|O$ object. And according
% to that version, this definition - and thus this statement -
% is (groundlessly) intended to be valid for any state vectors in~$\cal H$.
% (In general, not concerning about the domains of definitions of operators may lead to 
% 'paradoxes', as discussed in Ref.~\cite{gie}.)

\subsection{Features of the bra-ket scheme}

Efficient "working tools" provided by the bra-ket formalism \cite{dir1} include
a special designation $|\,\,\rangle$ for marking state vectors,
an "explicit" form of the projection-type operators, such 
as~$|u\rangle\langle v|$, and a convenient explicit representation 
of eigenstates of observables that form a complete set, such as
$|{\bf p}_1s_{z1},\ldots,{\bf p}_ns_{zn}\rangle$ 
 (where ${\bf p}_i$ and $s_{zi}$ denote
the momenta and projections of spins of particles).
 Matrix elements of operators between 
states belonging to a basis set    then take
a  convenient
form, such as~$\langle n_1',n_2',\ldots ,n_k'|O|n_1,n_2,\ldots ,n_k\rangle$.
(In our
scheme described below, all such tools are also available.)

Further, one needs to consider the drawbacks 
of the  above-described  
revised version of the bra-ket scheme. 
First, let us examine
the conventional bra-ket expressions
for matrix elements, such as $\langle u|O|v\rangle$.
The reasoning on this subject in this paragraph is basically
similar to the considerations by Gieres~\cite{gie}. 
The expressions $\langle u|O|v\rangle$
is introduced through
the equalities~(\re{1}). However, these equalities are not universally valid.
Indeed, let $D(O)$ and $D(O^\dag)$ denote
 the domains of definition of the operators $O$ and $O^\dag$.
The hermiticity relation~(\re{ad}), on which the 
equalities~(\re{1}) 
are based, applies only when $|v\rangle\in D(O)$ and
 \mbox{$|u\rangle\in D(O^\dag)$}   simultaneously.
 To provide a complete definition of the $O^\dag $ operator, the domain $D(O^\dag)$ 
 is to be specified.  
 It is to be accepted that 
 the (maximal) domain $D(O^\dag)$ consists of all the $u\equiv|u\rangle$
vectors for which the 
relation~(\re{ad}) takes place for any~$v\equiv|v\rangle$ vector belonging to~$D(O)$. 
Suppose that $u\notin D(O^\dag)$.
In this case, the above definition of the covector  $\langle u|O$ is unapplicable,
and 
the matrix element $\bigl(\langle u|O\bigr)|v\rangle$ entering Eq. (\re{1}) does not
exist. Suppose now that $v\notin D(O)$. Then the matrix element 
$\langle u|\bigl(O|v\rangle\bigr)$ that appears in  Eq. (\re{1}) does not
exist. In both cases, the expression $\langle u|O|v\rangle$ is undefined.  
Nevertheless,  the
matrix elements $(u,Ov)$ can exist even when $u\notin D(O^\dag)$ 
(assuming $D(O^\dag)\ne {\cal H}$)
  and the matrix elements $(O^\dag u,v)$ can exist even when  $v\notin D(O)$ 
  (assuming $D(O)\ne {\cal H}$).
  In such cases, the well-defined matrix elements $(u,Ov)$ and $(O^\dag u,v)$ 
  cannot be represented  by
 the  conventional bra-ket expression   
  $\langle u|O|v\rangle$.

 %And in these cases the bra-ket quantity
%$\langle u|O|v\rangle$ cannot represent the existing 
%$(u,Ov)$ and $(O^\dag u,v)$ matrix elements, as it should,
 %and thus it is poorly defined. 
 
% Thus the bra-ket matrix element
%$\langle u|O|v\rangle$ is well-defined only if, simultaneously, $|v\rangle\in D(O)$ 
%and $|u\rangle\in D(O^\dag)$. 

One cannot disregard such  matrix elements $(u,Ov)$ and $(O^\dag u,v)$.
Consider the case of operators $O$ that
can correspond to
observable quantities, specifically self-adjoint operators with unbounded spectra. In this
case, the expression  $\langle u|O|v\rangle$
is well-defined only if 
\mbox{$|u\rangle\in D(O)$}
 and \mbox{$|v\rangle\in D(O)$} simultaneously. However,  for any operator $O$ 
of this class,
there  exist, say, states~$|v\rangle$ that
do not belong to
$D(O)$. Indeed, when the spectrum of $O$ is discrete, writing 
\mbox{$\bigl(\langle v|O\bigr)\bigl(O|v\rangle\bigr)
=\sum_{\lambda,i}|\lambda\langle\lambda,i|v\rangle|^2$}, where $\lambda$ and 
$|\lambda,i\rangle$ are the
eigenvalues and eigenstates of the operator~$O$, one sees that there always exist 
states~$|v\rangle$ for which 
the series diverges,  indicating that the state $O|v\rangle$ does not exist. 
 When the spectrum of $O$ 
is continuous, a 
similar argument applies.\footnote{The more general Hellinger-Töplitz theorem 
(see, e.g., \cite{reed}) 
also states that  \mbox{$D(O)\ne {\cal H}$} always holds.}
 
   Thus,
   matrix elements $(Ou,v)$
  that cannot
  be represented in the  form $\langle u|O|v\rangle$ always exist for the 
  aforementioned  operators $O$.
For example, 
let $p_z$ be the operator of a  component of momentum. Suppose that for some~$n$, the  
 state $p_z^n |u\rangle$  is well-defined while the  state $p_z^n |v\rangle$ 
 is not well-defined.
  That is,
  the $u({\bf p})$ wave function
  decreases  sufficiently fast as $p_z$ increases, while  the 
  $v({\bf p})$ wave function does not.
 In this case,
 the well-defined  matrix element $\bigl(p_z^n u,v\bigr)$ 
 cannot be represented in the bra-ket form
 $\langle u|p_z^n|v\rangle$.  
 
Thus, instead of conventional bra-ket 
matrix elements of the  $\langle u|O|v\rangle$ type,   one, in general, should have
 used expressions like $\langle u|\bigl(O|v\rangle\bigr)$,
 $\bigl(\langle u|O\bigr)|v\rangle$, or  $\bigl(\langle u|O\bigr)\bigl(O'|v\rangle\bigr)$,
 which
 is overly
 cumbersome. 
%(One is  forced \cite{mes}
 %to handle such
 %expressions  also in all cases of 
% anti-linear operators, in particular, for the time reversion operator.) 
 
It was suggested \cite{gie} to adopt the rule
  that in expressions like $\langle u|O|v\rangle$, the operator acts only 
  to the right. However,  this rule does not extend to matrix elements such as
  $\bigl(\langle u|O\bigr)\bigl(O'|v\rangle\bigr)$. 
  Furthermore,  
  for expressions of the  $\bigl(\langle u|O\bigr)|v\rangle$ type, this rule
  is inconvenient, as  
  it requires replacing them 
  with their
  complex conjugate each time one encounters them.

Another drawback of the bra-ket scheme is that covectors of the   form $\langle u|O$
 entail potential
ambiguities regarding the domain of definition~\cite{gie}. 
To avoid this inconvenience, 
it has been suggested~\cite{gie,co} to represent the covector $\langle u|O$
with the expression $\langle O^\dag u|$ implying that 
$u\equiv|u\rangle$ in this context. However, in this case, the use of explicit
expressions for vectors
 specified by quantum numbers is precluded.

Herewith, we note that while the bra-ket formalism  incorporates adjoint operators 
as a constituent, it is possible to proceed without adjoint operators entirely in this context, as
 demonstrated  in the next 
section.\footnote{It is worth mentioning that, although this rarely occurs 
in practice, there may exist more than one operator that is adjoint to a given
operator, as F. Gieres mentioned to me. Therefore,
 relying on the one-to-one correspondence between an operator and its adjoint  
inherent in the
 bra-ket formalism
 is, in principle, also a limitation.}
 
In addition to all the issues mentioned above, 
 a significant disadvantage of the bra-ket formalism is
that this formalism, being a dual-space formalim, thus 
excludes
the natural one-space point of view.

There are known serious
 difficulties that students encounter with the bra-ket formalism.
In Ref. \cite{si},  for instance, the authors write in the Abstract: "We have been investigating 
the difficulties that the advanced undergraduate and graduate
 students have with Dirac notation.\ldots
 We find that many students struggle with Dirac notation, and they are not consistent 
 in using this
  notation". Probably, these difficulties might be alleviated if the one-space picture
  were employed to realize the representation-free
  approach. This picture would highlight the similarity of this 
  approach to
  simple vector algebra in its conventional one-space form,   
  as it is permanently utilized
throughout all branches of physics. 

Even though the bra-ket scheme is typically adopted as 
a representation-free formalism, it has faced criticisms in the literature. 
Gieres~\cite{gie} characterizes
 the bra-ket scheme as 
"representing a {\bf purely symbolic calculus}".
An utterance~\cite{gr}, also cited in~\cite{gie}, states: “Unfortunately, the 
elegance, outward clarity and strength of Dirac’s formalism
are gained at the expense of introducing mathematical fictions.\ldots One has a formal
‘machinery’ whose significance is impenetrable, especially for the beginner, and whose
problematics cannot be recognized by him". 
Also Steven Weinberg~\cite{wei} expresses 
his refusal to use the bra-ket formalism
speaking of its drawbacks.  
Instead, he employs the aforementioned 
representation-free,
mathematically oriented
formulation, while acknowledging
that it also has its disadvantages. 
  
\section{Present formulation}

First, we would like to  justify the necessity for a new formulation of the
representation free approach  concisely.
The mathematically oriented one-space formulation of this approach 
lacks practical calculation tools, such as those provided by the dual-space 
bra-ket formalism. Conversely,  
the latter formalism also has its drawbacks considered above. 
Gieres~\cite{gie} suggested that the representation-free approach itself is not that 
useful because the bra-ket
formalism does not adequately address the domains of definitions of operators. 
Furthermore, he argued that one should apply
the coordinate representation to a greater extent, as these domains of 
definitions are naturally determined  in this representation. 

However,  due to the significant advantages
of the representation-free approach when working
with general relations, a much better way out would be to  
construct a realization of this approach that is more suitable than the 
bra-ket formalism in the aforementioned and other respects. While, 
to obtain information regarding the specific domains of definitions of the involved operators,
one could utilize suitable representations before transitioning to a 
 representation-free description.

An appropriate realization of the representation-free approach should 
incorporate all the tools of the type of those in Dirac's formalism, that
 facilitate practical calculations. It should also utilize a 
 one-space description
  while allowing for a dual-space interpretation simultaneously. The 
  dual-space description is needed
  for specific noteworthy purposes, such as  rigged Hilbert
   space \cite{gel} applications \cite{gie,bo,oh,oh1} or refined interpretations
   of quantum theory \cite{gan, ott}. Additionally,
   a dual-space interpretation would align more closely  
    with the widely used bra-ket formalism. The representation-free scheme 
    constructed here meets all these requirements. 
    
%(In various expressions below, state vectors may be the generalized ones.)

 %A  representation-free scheme suggested below 
%contains 
%all the tools    needed to operate in quantum mechanics that are inherent in
%the bra-ket scheme. 
%At the same time,
%it is free of all the above listed  drawbacks of the bra-ket  scheme.
%The relation between the present scheme and the bra-ket scheme is very simple and,
%in particular, the bra-ket notation may be introduced in a simple way proceeding
%from the present approach. 
Let us move on to the present representation-free scheme.
As the bra-ket formalism, the present scheme includes a special marking of state vectors. 
We denote state vectors specified by quantum numbers as 
$/{\bf p}_1s_{z1},\ldots,{\bf p}_ns_{zn}/$ (and, for  stationary scattering 
states~\cite{tay}, as
\mbox{$/{\bf p}_1s_{z1},{\bf p}_2s_{z2}\pm/$}). We  also
mark state vectors as~$/u/$  in certain other cases, but we simply
use symbols like $u$ when this does not lead to confusion.

For the  quantum mechanical scalar product, we  suggest 
the expression like~$u\lcdot v$ or, for instance, $/{\bf r}/\lcdot/plm/$, or 
$/{\bf p}_1,\ldots,{\bf p}_n/\lcdot v$. 
While in the  bra-ket scalar product expression
$\langle u|v\rangle$  the bra vector $\langle u|$ is 
a linear functional,
 in the present $u\lcdot v$ expression
  $u$ is a vector. Thus, $u$ and~$v$ belong to the same space.
The $u\lcdot v$ expression
is an analog to the expression for the scalar product of
vectors in finite-dimensional spaces, such as  ${\bf u}\cdot {\bf v}$.
(Of course, 
 $v\lcdot u=(u\lcdot v)^*$, $v\lcdot v\ge0$, and $u\lcdot v$
 is linear
 in $v$ and 
 thus anti-inear in~$u$.) The adjoint operator $O^\dag$ 
  of a linear  operator $O$ is defined in the conventional way: 
 \be O^\dag u\lcdot v
 =u\lcdot Ov\la{3}\ee
 where the domain of definition of $O^\dag$ is indicated 
 in relation to Eq.~(\re{ad}).
 (If $O$ is 
 anti-linear then 
 $O^\dag u\lcdot v=Ov\lcdot u$.) Eq. (\re{3}) is simpler than Eq.~(\re{ad}).
 Unlike the bra-ket formalism,
 adjoint operators are not a component of the present scheme. For this 
 reason and others, 
 this scheme is less involved than the bra-ket one.

As discussed in the previous section, 
the conventional bra-ket expressions for matrix elements, such as~$\langle u|O|v\rangle$
or $\langle u|OO'|v\rangle$,
can be poorly defined for certain states $|u\rangle$ or
$|v\rangle$.  Matrix elements, that do exist and include such states
cannot be represented by bra-ket expressions.  
And the corresponding well-defined expressions 
%that could
%represent such matrix elements 
are overly cumbersome.  
This drawback is, of course, absent here. In the present formulation,
all possible matrix elements are represented by
simpler expressions such as 
\mbox{$u\lcdot Ov$} or \mbox{$Ou\lcdot v$}, or \mbox{$Ou\lcdot O'v$}. 
In contrast to the bra-ket representation of matrix elements, 
these expressions clearly demonstrate 
the needed correspondence between  the domains of definition of  
operators and states they act upon.{\footnote{In LaTex, 
to produce the large-sized 
 dot we employ the command {\tt \symbol{"5C}boldsymbol}\{{\tt \symbol{"5C}cdot}\}.}
%\footnote{In LaTex, 
%to produce the large-sized 
% dot we employ,
% one can define a
%command called, for example,
%{\tt \symbol{"5C}lcdot}. Here's how one can do it:
%\mbox{{\tt 
%\symbol{"5C}newcommand\{\symbol{"5C}lcdot\}\!\!\!
%\{\symbol{"5C}mathbin\{\symbol{"5C}stackrel\{\symbol{"5C}centerdot\}\{\}\}\}}}.}
 
In some (though not all) cases, it is convenient to omit the dot in the 
scalar products of the form \mbox{$/u/\lcdot O/v/$}  
writing them simply as $/u/ O/v/$. (One has $/u/ O/v/^*=O/v/\lcdot/u/$.)
Matrix elements of an operator in a basis are then represented like 
\[/n_1',n_2',\ldots n_k'/O/n_1,n_2,\ldots n_k/.\]
In the case of matrix elements 
 $/\ldots j_1m_1/O_{jm}/\ldots j_2m_2/$
 of an irreducible tensor operator~$O_{jm}$ between states with specified angular momenta, 
 it is natural to denote the corresponding reduced matrix elements 
 as~$/\ldots j_1//O_j//\ldots j_2/$. 

We introduce multiplication signs $*$ between constituents
of emerging expressions, such as
 scalar products or state vectors and scalar products.
This appears as $u*v\lcdot w$, or
 $u\lcdot v*u'\lcdot v'*w$. 
The multiplication signs provide clarity, and in particular,  they 
are required for what
follows.

One should distinguish between  the
expressions   $cu\lcdot v$ and $(cu)\lcdot v$, where $c$ is a constant.
While the second expression represents
the scalar product of the states $cu$ and $v$, the first one 
 is interpreted as the product of the constant $c$ and the scalar product $u\lcdot v$
 (unlike expressions 
such as $Ou\lcdot v$).
One has $(\sum_ic_iu_i)\lcdot v=\sum_i(c_iu_i)\lcdot v=\sum_ic_i^*u_i\lcdot v$. 
If, for example, in the above relations the constants 
$c$ or $c_i$ themselves are scalar products then for clarity the replacement $c$ by $c*$ or
$c_i$ by $c_i*$ and $c_i^*$ by $c_i^**$ is to be done.

Further,  as in the case of the bra-ket formalism, we propose an 'explicit' expression 
for projection-type operators. In the present scheme, this expression is of the 
form~$u*v\,\lcdot$.  
It incorporates the multiplication sign between the vectors $u$ and $v$ while 
the dot now
follows the vector~$v$.  
The result of the action of the operator ~$u* v\,\lcdot$ 
on a state~$w$ is then expressed as~$u*v\lcdot w$.
Here, $v\lcdot w$ represents the scalar product as defined above.
Thus,  the action of projection-type operators 
is self-explanatory. 
It follows from Eq.~(\re{3}) that
 $(u*v\,\lcdot)^\dag=v*u\,\lcdot$. 

Let us write down some simple relations employing the above expression for projection-type
 operators. 
 If $\{/m/\}$ is an orthonormal basis, then 
 the identity operator $I$ can be expressed as $I=\sum_{m}/m/*/m/\lcdot$. (Of course, 
 this expression is equivalent to
 the expansions of state vectors over the set~$/m/$, such as
 \mbox{$u=\sum_m  /m/*c_m$}, where $c_m=/m/\lcdot u$.)
Similarly,  if $/\nu/$  is, e.g., a complete set of continuous  observables
%\mbox{($/\nu/\lcdot/\nu'/=\delta(\nu-\nu')$)}, 
then $I$ can be written as 
\mbox{$\int d\nu /\nu/*/\nu/\lcdot$}. 
 
Inserting projection operators into scalar products can, of course,  be helpful.
For example, using the latter expression for $I$, one obtains
\[/q/\lcdot\psi\equiv /q/\lcdot I\psi=\int d\nu/q/\lcdot/\nu/*/\nu/\lcdot\psi.\]
In particular, the $q$-representation wave function is expressed in terms of 
the $\nu$-representation wave function
in this way.

 Let $\{/m/\}$ and $\{/n/\}$ be 
 orthonormal bases. One may formally represent any operator~$O$ 
 in terms of the projection-type
 operators $/m/*/n/\lcdot$:
 \[O=
\sum_{m,n}\,/m/*/m/\lcdot O /n/*/n/\lcdot\equiv /m/*/m/O /n/*/n/\lcdot\equiv
/m/ O /n/*/m/*/n/\lcdot.\]
In particular, one may write the identity operator $I$ as
\mbox{$\sum_{m,n}\, /m/\lcdot/n/*/m/*/n/\lcdot$}, which reduces to
the  expression for $I$ written above
if the bases $\{/m/\}$ and $\{/n/\}$ coincide.

If $O_{pt}$ is a projection-type operator  
then the product $O_1*O_{pt}*O_2$ of three operators can be formally  represented
as a single projection-type operator. Denote this latter operator as 
$/\mu/*/\nu/\lcdot$ and $O_{pt}$ as $/m/*/n/\lcdot$. Then  we have
\mbox{$/\mu/=O_1/m/$} and
$/\nu/=O_2^\dag/n/$. 

One may write the product of projection-type operators  as such an operator times
the product of scalar products, 
\mbox{$\prod_{i=1}^nu_i*v_i\,\lcdot=
\left(\prod_{i=1}^{n-1}v_i\lcdot u_{i+1}\right)u_1*v_n\lcdot$.} 
%Let $O_i$ be some
%operators, so that $u_i*v_i\lcdot O_i$ is the product of two operators. 
%One has a generalization 
%of the above relation, 
%\[\prod\nolimits_{i=1}^nu_i*v_i\lcdot O_i=
%\left(\prod\nolimits_{i=1}^{n-1}v_i\lcdot O_iu_{i+1}\right)u_1*v_n\lcdot O_n.\]
All this shows the utility of the explicit expression for projection-type operators.

Up to this point, we formulated the present scheme as a one-space scheme. 
Let us incorporate the dual-space point of view in the present scheme.
To this aim, we will reinterpret our notation.
 In the above scalar product
expression~\mbox{$u\lcdot v$},  $u$ was considered a 
state vector,
and the dot represented the scalar product sign.
And now we will treat $u\,\lcdot$ as an indivisible
symbol. Namely, we will
consider $u\,\lcdot$ as the other designation for the linear functional 
 $G_u$ in the relation \mbox{$G_u(v)=u\lcdot v$}. That is, one will understand
 $u\lcdot v$ as 
 $u\lcdot(v)$ under this interpretation. Consequently, we will interpret  
 $u\,\lcdot$ as a covector. This interpretation was proposed in the 
 preliminary version of the present paper \cite{efr}.\footnote{In particular, 
  it has been 
 pointed out in \cite{efr} (since the 2023 [v7] version)
 that an easy way to learn the bra-ket formalism is to proceed from this interpretation,
 as described below.
 A similar approach is also advocated in the recent paper \cite{lev}.} 
 (The notation $Ou\,\lcdot$ has 
 the same meaning as $(Ou)\,\lcdot $. In contrast, if $c$ is a constant then $cu\lcdot$ is
 interpreted as the product of the constant $c$ and the covector $u\lcdot$.
  If $c_i$ are constants and $u_i$ are state vectors, then we have
 \mbox{$(\sum_ic_iu_i)\lcdot=\sum_ic_i^*u_i\lcdot$}.
 If, for example, $c_i$ and $c_i^*$
 are scalar products, then for clarity they should be replaced
 by $c_i*$ and $c_i^**$.)

In this way, one shifts to  
the dual-space description discussed in the previous section regarding the bra-ket formalism. 
According to this description,  one deals with the spaces of state vectors and 
state 
covectors. The scalar product $u\lcdot v$ becomes
 the "inner product" of a state covector $u\,\lcdot$ and a state vector $v$. 
 And in both one-space and dual-space
  versions of the present scheme, the projection-type 
  operator $u*v\,\lcdot$ is the "outer product" of a state vector $u$ and 
  a state covector $v\,\lcdot$.  
  
Thus, in contrast to the  bra-ket, dual-space, formalism, 
the present formulation allows both the one-space
and dual-space interpretations.

As mentioned in the previous section, the bra-ket expressions for covectors, such as
$\langle\psi|O$ or $\langle\psi|O^\dag$,  entail potential ambiguities regarding
the domain of definition. This drawback is absent here. 
Indeed, for instance,
the  covector $\langle\psi|O^\dag$ is expressed  as $O\psi\,\lcdot$ in the present formulation.
From this expression, it is apparent that $\psi$ should belong to the domain of 
definition of the operator $O$.

(One can express a mapping  $\psi\lcdot\rightarrow O\psi\lcdot$ of covectors 
in terms of an operator acting on the covectors $\psi\lcdot$. Let us denote this operator 
as $\bar O$ and  position
it to the right of covectors. Then, we have $O\psi\lcdot\equiv\psi\lcdot{\bar O}$.
Accordingly, scalar products of the form
$O\psi\lcdot\chi$ can  be rewritten as $(\psi\lcdot{\bar O}) \chi$.
If the operator $O$ is linear, then the corresponding operator $\bar O$ is also linear.
In terms of $\bar O$, the relation defining conjugate operators takes the form 
 $(\psi\lcdot{\bar O}) \chi=\psi\lcdot O^\dag\chi$. If
this relation holds for any states $\psi$ and $\chi$, 
one may view the operator $\bar O$ as the operator $O^\dag$ acting to the left, i.e., on
covectors. Accordingly, in this case, one can denote $\bar O$ as $O^\dag$,
 assuming  
that in the expression $\psi\lcdot O^\dag\chi$
the operator $O^\dag$ may act both to the right and to the left. This applies, in particular,
in the finite-dimensional case.)

According to the above,
the correspondence between the bra-ket notation and the notation 
of the dual-space version
of the present scheme is as follows,
\[|\psi\rangle\Leftrightarrow \psi,\qquad\langle\psi|\Leftrightarrow \psi\,\lcdot,\qquad
\langle\psi|O\Leftrightarrow O^\dag \psi\lcdot.\]
Further, when transitioning from the bra-ket formulation to the present formulation,
the matrix element expression $\langle\psi|O|\chi\rangle$ should be replaced 
  with either 
of the expressions $\psi\lcdot O\chi$ and  $O^\dag \psi\lcdot \chi$,
provided that each 
of the latter expressions is well-defined. 
 And if one of these expressions is not well-defined,  
then the  bra-ket expression $\langle\psi|O|\chi\rangle$ 
has no sense, as discussed in Sec. 2 C. 
An easy opportunity 
to master the widely used bra-ket formalism is to start with the one-space 
version of the present scheme and  then move on
 to its dual-space version, utilizing
the listed
 correspondence.

\section{Conclusion}

To carry out representation-free considerations in quantum theory,
 the bra-ket 
formalism 
is generally employed. While it offers 
helpful tools for this purpose, it also has inherent drawbacks.
Some of them arise from the fact that the bra-ket scheme
is not well-suited  for handling the domains of definitions of 
 operators involved. One more drawback is that
 the bra-ket formalism, as a dual-space formalism, excludes the 
 natural one-space point of view. Sec. 2 provides 
 an analysis of the bra-ket formalism and addresses its drawbacks.
 
 The present work gives a formulation of the representation-free approach 
which  
is entirely free of the 
 drawbacks of the bra-ket formalism. 
 In particular, unlike the bra-ket formalism,
 the present scheme 
 allows both the needed one-space and 
 dual-space interpretations. With all that,
 the present formulation provides 
efficient means for conducting representation-free
considerations, akin to all such means 
  offered by
 the bra-ket formalism. Herewith, the present scheme is less
 involved  than the bra-ket one.
 The constructed "universal" scheme is very well-suited for performing representation-free
  practical
 calculations. 
 
Moving on from
 the bra-ket formalism to the present formulation is easy.
And an easy opportunity 
to master the widely used bra-ket formalism would be
to proceed from the present formulation.

 \section{Acknowledgements}
    The author is grateful to A. Aiello and A. L. Barabanov, whose comments helped to
    improve the presentation. The author is indebted
     to F. Gieres for his helpful feedback regarding Ref.~\cite{gie}.
     
 %  \section{Statements and Declarations} 
  % The author complies with Ethical Standards as presented in Instruction for Authors 
  % of the journal. 
 %No funding was received to assist with the preparation of the manuscript. 
 %The author has no conflicts of interest to declare that are relevant to the 
 %content of this paper. 
 
% \pagebreak


\begin{thebibliography}{99}
\bibitem{dir1} P A M Dirac  
{\it 
The Principles of Quantum Mechanics
} 
4th edn 
  (Oxford: 
Clarendon Press) Ch 1, p 18; Ch 2, p 23, 26 (1958)
\bibitem{dir2} P A M Dirac  
%:  A New Notation for Quantum Mechanics. 
{\it 
Math. Proc. 
of the Cambridge Phil. Soc.
} 
{\bf 35} 416 (1939)
\bibitem{gie} F Gieres  
%: Mathematical surprises and Dirac's formalism in 
%quantum mechanics.
{\it 
Rep. Prog. Phys.
} 
{\bf 63} 1893  (2000)
%Doi 10.1088/0034-4885/63/12/201
\bibitem{ja} J M Jauch   
%: On Bras and Kets. 
%in  
{\it 
Aspects of Quantum Theory
} 
(eds) A Salam and E P Wigner 
(Cambridge: University Press) p 137 (1972; 2010)
\bibitem{gr} D Grau  
{\it  
\"Ubungsaufgaben zur Quantentheorie - Quantentheoretische
 Grundlagen
 } %3 edn
 (M\"unchen: C. Hanser Verlag) (1993)
\bibitem{wei} S Weinberg  
{\it  
Lectures on Quantum Mechanics
} 
(Cambridge: University Press) 2nd edn. Preface, p. xviii; Ch 3, Sec 3.3., p. 65, 71
 (2015)
%\bibitem{medv} B.V. Medvedev,  {\it Fundamentals of Theoretical Physics} 
%(Fizmatlit, 2007)
\bibitem{si} C  Singh  and E Marshman  
%: Investigating Student Difficulties with Dirac Notation.
 {\it arXiv}: 1510.01296 (2015); %in %{\it 
 {\it PERC Proceedings}
 %}, 
 (eds) 
P V Engelhardt, A D Churukian and  D L Jones (Portland: PERC)
p. 345 (2013)%; doi:10.1119/perc.2013.pr.074 
 % (eds.)  , pp.
 %345-348. 
 %doi:10.1119/perc.2013.pr.074
 \bibitem{gel} I M Gel’fand  and N Ya Vilenkin   
 {\it 
 Generalized Functions. Vol. 4 - Applications
of Harmonic Analysis
} 
(Chelsea: Am. Math. Soc.) (2016)
 \bibitem{bo} N N Bogolubov, A A  Logunov  and  I T Todorov   
 {\it 
 Introduction to 
Axiomatic Quantum Field Theory
}
(San Francisko: Benjamin Cummings) Ch 1, Secs 1.5, 1.6 (1975)
\bibitem{oh} S Ohmori {\it J. Math. Phys.} {\bf 65} 123502 (2024)
\bibitem{oh1} S Ohmori and J Takahashi {\it J. Math. Phys.} {\bf 63} 123503 (2022)
%\bibitem{be}  F A Berezin  and M A Shubin   
%{\it 
%The Schr\"odinger Equation
%} Mathematics and
%its Applications Vol. 66
%(Dordrecht: Kluwer) (1991)
\bibitem{gan} S V Gantsevich {\it Ann. Math. Phys.} {\bf 5} 196 (2022)
\bibitem{ott} H C \"Ottinger {\it J. Phys. Commun.} {\bf 8} 105004 (2024)
\bibitem{mes} A Messiah   
{\it 
Quantum Mechanics
}% 4 edn
(New York: Dover) Ch VII, Sec I (2020)
\bibitem{co} C Cohen-Tannoudji, B Diu  and F Lalo\"e   
{\it 
Quantum Mechanics}  
(Paris: Wiley
 and Hermann) {\bf Vol. 1}, Ch II, Sec B (1977)
 \bibitem{reed} M Reed  and  B Simon  
{\it 
Methods of Modern Mathematical Physics. Vol. 1 - Functional Analysis
} %2 edn
(San Diego: Academic) Ch II, p 38, 43; Ch III, p 84 (1980)
% \bibitem{fa} G Fano 
 %{\it  
 %Mathematical Methods of Quantum Mechanics 
 %} 
 %(New York: McGraw-Hill) (1971)
%\bibitem{ha} P R Halmos, {\it A Hilbert space problem book} (Springer, Heidelberg, 1982).
%\bibitem{ma} E Marshman and C Singh {\it Eur. J. Phys} {\bf 39} 015707 (2018); 
%{\it PERC Proceedings}
 %}, 
% (eds) S Wolf, M Benett and  B Frank p. 309 (2020)
%\bibitem{lan} L D Landau and E M Lifshits {\it Course of Theoretical Physics. Vol. 3 - Quantum
% Mechanics. Non-relativistic theory} 3rd edn. (New York: Pergamon) (1991)
\bibitem{tay}  J R Taylor  
 {\it 
 Scattering Theory: The Quantum Theory of 
 Nonrelativistic Collisions
 } %5 edn
 (New York: Dover) Ch 10, Sec 1 (2012)   
\bibitem{efr} V D Efros  {\it arXiv}: 2212.10597 (2022); {\it ResearchGate preprint} (2021)
\bibitem{lev} J Levy and C Singh {\it Am. J. Phys.} {\bf 93} 46 (2025)
 \end{thebibliography}
\end{document}